\pdfoutput=1
\documentclass[11pt]{article}

\usepackage[final]{acl}

\usepackage{times}
\usepackage{latexsym}

\usepackage[T1]{fontenc}
\usepackage[utf8]{inputenc}

\usepackage{microtype}

\usepackage{inconsolata}

\usepackage{graphicx}

\usepackage{amsfonts,amsmath,amssymb,amsthm} % for math;

\usepackage{amsfonts,amsmath,amssymb,amsthm} % for math;
\usepackage{amsmath} % needed for align* environment
\usepackage{tabularx}
\usepackage{multirow}
\usepackage{booktabs}
\usepackage{paralist}
\usepackage{microtype}
\usepackage{caption} % customize table captions
\usepackage{array} % for column width adjustments
\usepackage{rotating} % for rotating tables
\usepackage{svg}
\usepackage{xcolor}
\usepackage{inconsolata}
\usepackage{booktabs} 
\usepackage{amsfonts}
\usepackage{amssymb}
\allowdisplaybreaks
\usepackage{tabularray}
\usepackage{tabularx}
\usepackage{multirow}
\usepackage{arydshln}
\usepackage{CJKutf8}
\usepackage{authblk}
\usepackage{wrapfig}
\usepackage{enumitem}
\usepackage{xcolor}
\usepackage{float}

\usepackage{hyperref}
\usepackage{graphicx} % For including graphics
\usepackage{subcaption} % For subfigures

\usepackage{stfloats}

\newcolumntype{L}{>{\RaggedRight\hangafter=1\hangindent=0em}X}
\title{ActorMind: Emulating Human Actor Reasoning for Speech Role-Playing}

\author{Xi Chen,  Wei Xue \thanks{Corresponding author}, Yike Guo \\
        The Hong Kong University of Science and Technology\\
        \texttt{chenxi.mail.1005@gmail.com}, \texttt{weixue@ust.hk}}

\begin{document}
\maketitle

\begin{abstract}

Role-playing has garnered rising attention as it provides a strong foundation for human-machine interaction and facilitates sociological research. 
However, current work is confined to textual modalities, neglecting speech, which plays a predominant role in daily life, thus limiting genuine role-playing. 
To bridge this gap, we conceptualize and benchmark speech role-playing through \textbf{ActorMindBench}, and we present a corresponding reasoning framework, called \textbf{ActorMind}. 
Specifically,
(1) \textbf{Speech Role-Playing} enables models to deliver spontaneous responses with personalized verbal traits based on their role, the scene, and spoken dialogue.
(2) \textbf{ActorMindBench} is a hierarchical benchmark comprises \textit{Utterance-Level content} with 7,653 utterances, \textit{Scene-Level content} with 313 scenes, and \textit{Role-Level content} with 6 roles.
(3) \textbf{ActorMind} is an off-the-shelf, multi-agent chain-of-though style reasoning framework that emulates how human actors perform in theaters.
Concretely, ActorMind first reads its assigned role description via \textbf{\textcolor{olive}{Eye Agent}}, then comprehends emotional cues within contextual spoken dialogues through \textbf{\textcolor{cyan}{Ear Agent}.} 
Subsequently, \textbf{\textcolor{violet}{Brain Agent}} generates a descriptive emotional state, and finally, \textbf{\textcolor{orange}{Mouth Agent}} delivers the scripts infused with corresponding emotion state. 
Experimental results demonstrate the effectiveness of ActorMind in enhancing speech role-playing. 
The project page is available at \url{https://github.com/OzymandiasChen/ActorMind}.

\end{abstract}

\section{Introduction}

Role-playing (RP) involves customizing models, particularly Large Language Models (LLMs), to generate spontaneous, human-like responses with personalized traits 
based on the surrounding context 
\citet{rpallmsurvey}. 
Recently, RP has garnered rising attention as it represents genuine machine intelligence and creativity. 
It enables LLMs to 
offer nuanced interaction experiences for users \citep{rolellm}, 
provide emotional value \citep{characterllm, johansson2025open}, and support sociological research \citep{mmrole, chateval, zhang-etal-2025-sotopia}.

Numerous benchmarks and methods \citep{characterllm, rolellm, zhang2025revealing} have been proposed recently. 
However, they primarily focus on text modality, overlooking that human activities occur across multiple modalities, including text, audio, and vision. 
Among these, audio, especially speech, which is predominant for conveying emotions and attitudes in daily life, reveals persona in a direct and vivid way. 
Both Large Language-Audio Models (LLAMs) and Text-to-Speech Synthesis (TTS) models are capable of generating speech: LLAMs \citep{qwenomni, gpt4o} exhibit strong capabilities in instruction-following, while recent TTS models enable fast-speed \citep{valle2} and zero-shot speech synthesis \citep{cosyvoice}. 
Despite these advances, existing models still lack the ability to produce spontaneous, persona-consistent speech responses.
Therefore, developing publicly available benchmarks and principled reasoning frameworks for speech role-playing is crucial.

To bridge this gap, 
we 
(1) conceptualize speech role-playing; 
(2) propose a publicly available benchmark ActorMindBench, along with corresponding tool pipeline;
% for future benchmark expansion; 
and (3) introduce ActorMind, a multi-agent \citep{llmagentsurvey2025} chain-of-thought (CoT) style \citep{cot} speech role-playing method inspired by emulating the script delivery process of human actors in theaters \citep{actor}.

\textbf{Speech Role-Playing} involves injecting roles into speech generation and interacting via delivering target scripts with personalized verbal attributes, such as \textit{“Wistful flirtation, tinged with a hint of playful vulnerability”}, based on the context, including scene descriptions and historical spoken dialogues.

\textbf{ActorMindBench} is hierarchically designed with three levels of data.
\textit{Utterance-Level} includes speech segments with text content and speaker labels; \textit{Scene-Level} includes scene boundaries and descriptions; \textit{Role-Level} includes role profiles.
Specifically, ActorMindBench is constructed from well-known television sitcoms, ensuring the authenticity and naturalness of the data. 

\textbf{ActorMind} is a multi-agent CoT style reasoning framework that facilitates speech role-playing by emulating how human actors perform in theaters. 
Typically, before performing, human actors first read the scripts to understand their roles and gain a rough understanding of how the scene develops. 
While acting, they carefully listen to the tones and emotions conveyed by other actors. 
By combining their character, scene description, and the emotions of others, they formulate their own emotion and tone for delivering the next line. 
Finally, they deliver the line spontaneously \citep{actor}.
Inspired by this process, ActorMind, shown in Figure \ref{fig:AM_overview}, conceptualizes four agents—Eye, Ear, Mouth, and Brain. 
The \textbf{\textcolor{olive}{Eye Agent}} handles character profile and scene script reading; 
the \textbf{\textcolor{cyan}{Ear Agent}} focuses on listening to the speech tones of others; 
the \textbf{\textcolor{violet}{Brain Agent}} is responsible for emotion state reasoning; 
and, aided by Retrieval Argument Generation (RAG) \citep{ragsurvey, zhang2026biminddualheadreasoningmodel}, the \textbf{\textcolor{orange}{Mouth Agent}} delivers the script with the desired emotion and voice by referencing emotionally similar historical speeches.
Experiments on ActorMindBench validate the effectiveness of ActorMind. 
Notably, ActorMind is an off-the-shelf reasoning framework that can be easily utilized.

Briefly, our contributions are threefold:

\begin{enumerate}
[nolistsep,leftmargin=*]
  \item We propose ActorMindBench, a publicly available, hierarchical benchmark for speech role-playing, along with its construction pipeline.
  It includes \textit{Utterance-Level content} with 7653 utterances, \textit{Scene-Level content} with 313 scenes, and \textit{Role-Level content} with 6 roles.
  \item We introduce ActorMind, a multi-agent CoT style, off-the-shelf speech role-playing method inspired by how human actors perform in theaters.
  \item The evaluation results provide compelling evidence of ActorMind’s remarkable performance.
\end{enumerate}

\section{Related Works}

\subsection{Role-Playing in LLM}
The advancement of LLMs \citep{bert, gpt4, llama3} has significantly shaped and catalyzed the development of role-playing. 
By leveraging supervised fine-tuning \citep{sft, chang2026balora, chang2025lora} and in-context learning \citep{icl, li2025m}, role-playing can be achieved by training \citep{pcgr2025naacl, ssl2022mmasia} or prompting LLMs with high-quality, character-specific dialogues.
The majority of existing work focuses on the text modality. 
For example, \citep{chen2022large} is built upon the well-known Harry Potter universe to establish authentic role-playing, while \citep{rolellm} is developed using artificial datasets, enabling role-playing agents across a wide range of environments and circumstances.
Furthermore, \citep{mmrole} is the first work dedicated to role-playing in the language-vision modality, extending the boundaries of role-playing into the multimodal domain.

In this work, we further extend role-playing into the speech domain, as speech is the predominant modality for conveying emotio and information \citep{mmrbn2024icassp}.
Specifically, ActorMindBench (Section~\ref{sec:actormindbench}) and ActorMind (Section~\ref{sec:actormind}) together provide a comprehensive benchmark and a principled reasoning framework for speech role-playing.

\subsection{Speech Generation Models}

Both Large Language–Audio Models (LLAMs) and Text-to-Speech (TTS) models are capable of generating speech, yet they exhibit complementary strengths and limitations with respect to speech role-playing.
(1).
\textbf{LLAMs} \citep{qwenomni, gpt4o} are designed to perform complex reasoning and instruction following, with inputs and outputs spanning both text and audio modalities. 
Representative models such as Qwen-Omni \citep{qwenomni} and GPT-4o \citep{gpt4o} demonstrate strong capabilities in multimodal understanding. 
However, their supported voice inventories are typically very limited, often ranging from only a few to around ten voices. 
This constraint fundamentally restricts their ability to perform fine-grained role-playing, such as convincingly portraying specific characters (e.g., “Harry Potter”).
% \textbf{(2).} 
(2).
\textbf{TTS models}, such as SparkTTS and IndexTTS \citep{sparktts, indextts}, take text as input and generate corresponding speech. 
These models exhibit strong in-context learning and zero-shot capabilities for voice cloning and speaking style transfer. 
Nevertheless, they generally lack role-playing abilities: they struggle to adopt role-specific speaking styles and to respond spontaneously and coherently to dynamic scenes and dialogues.

In role-playing with LLMs, generated responses typically exhibit strong textual persona traits, such as characteristic phrasing or catchphrases \citep{ma2026stable}.
Analogously, speech role-playing requires generated speech to convey spontaneous and authentic character traits—for example, a speaking style described as “wistful flirtation, tinged with a hint of playful vulnerability.”
In this work, ActorMind equips speech generation models with such capabilities, serving as a generalizable framework for speech role-playing.

\subsection{Chain-of-Thought Style Reasoning}

Chain-of-thought reasoning \citep{cot, ling2026neural} is a technique in which models are guided to generate explicit intermediate reasoning steps, enabling more effective handling of complex problems that require multi-step inference. It is based on the assumption that generating more tokens for reasoning can lead to improved performance \citep{muennighoff2025s1, jin2026inference}.
CoT prompting has advanced numerous fields, including multilingual factual reasoning \citep{Weihua_Huang_Liu_Vangani_Zou_Tao_Wu_Aw_Chen_Lee_2026}, mathematical reasoning \citep{jiang2025drp}, and abstract summarization \citep{yuan2025understanding}. It has also been shown to mitigate hallucinations \citep{weihua-etal-2025-ccl} and extend to multimodal domains such as autonomous driving \citep{zeng2025futuresightdrive}.

In this work, we extend CoT to the speech role-playing domain. By emulating human actors performing in theater, ActorMind adopts an “eye-ear-brain-mouth” reasoning process, enabling intuitive and coherent speech-based role-playing.

\subsection{LLM Agent}

An LLM agent \citep{mohammadi2025evaluation} is a computational system that leverages a LLM as its core reasoning engine, enabling it to interpret instructions, make decisions \citep{jiang2026scribestructuredmidlevelsupervision, zhu2026symphony}, and interact with external tools \citep{yang2026evotool, yang2026tooltree} or environments \citep{yao2022react, jiang2026magma} to accomplish complex tasks.
LLM agents have driven progress across numerous fields, including social network simulation \citep{zhang-etal-2025-ga} and logical reasoning \citep{zhang2026logical}.

In this work, the proposed “eye-ear-brain-mouth” reasoning process is realized through a set of coordinated agents. The Ear Agent is powered by Automatic Speech Recognition and Speech Emotion Captioning, enabling it to perceive both linguistic and emotional signals. The Brain Agent, powered by an LLM, performs high-level emotion state reasoning. Finally, the Mouth Agent leverages RAG to deliver scripts infused with corresponding emotion state.

\section{ActorMindBench}
\label{sec:actormindbench}

ActorMindBench is hierarchically designed with three levels of data: \textit{Utterance-Level}, \textit{Scene-Level}, and \textit{Role-Level}. 
An example from ActorMindBench is shown in Figure \ref{fig:AMB_example} in Appendix \ref{appendix:AMB_exampledata}.
In this section, we will present our design principle and construction pipeline.

\begin{figure*}[htbp]
  \centering
   \includegraphics[width=.7\linewidth]{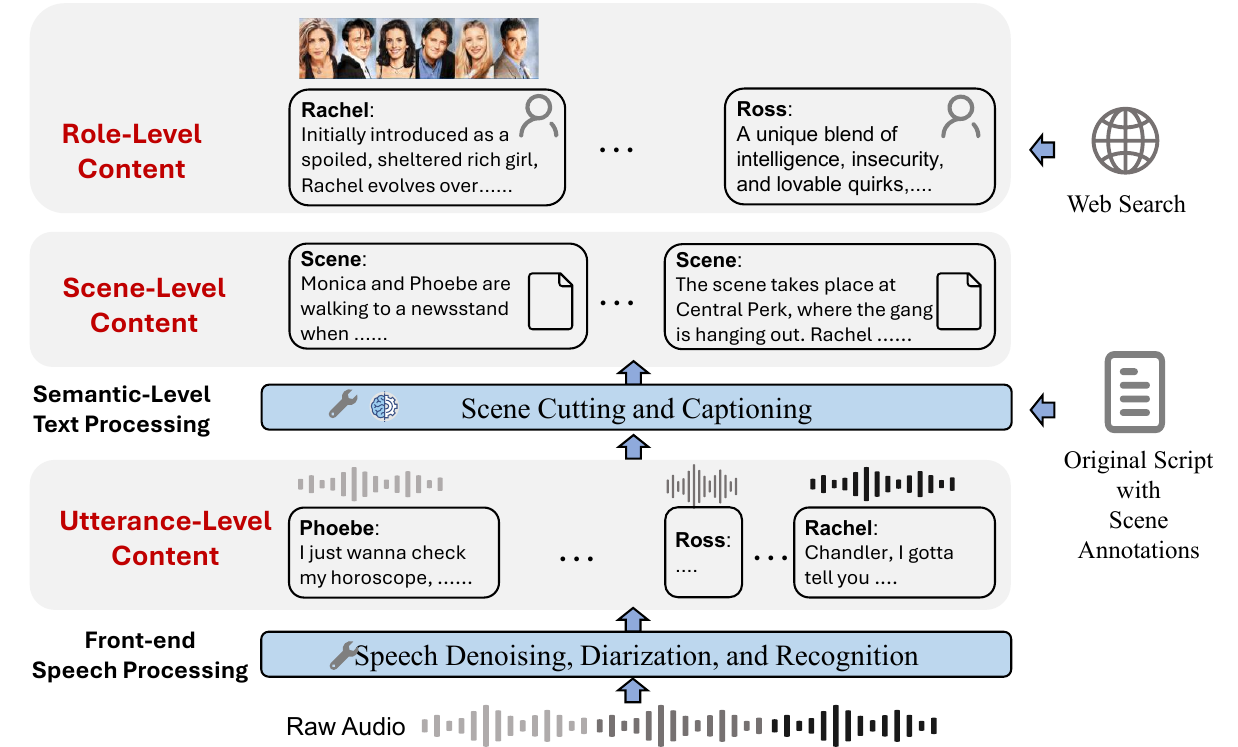}
  \caption{
  Overview of ActorMindBench.
  ActorMindBench comprises three content levels: \textit{Utterance-Level} includes speech segments with text content and speaker labels; \textit{Scene-Level} includes scene boundaries and descriptions; \textit{Role-Level} includes role profiles.
  }
  \label{fig:dataset}
\end{figure*}

\subsection{Design Principle}
\label{sec:AMB_design_principle}

\paragraph{Components.}
Human beings naturally enjoy role-playing and theatrical performance, which motivates the design of role-oriented benchmarks and modeling methods. 
Inspired by \citet{baumol1965performing}, ActorMindBench is structured around three levels of content:
\begin{itemize}[noitemsep,nolistsep,leftmargin=1em]
    \item \textit{Utterance-Level}: individual lines from theater scripts, including the speaker name, textual content, and corresponding speech data;
    \item \textit{Scene-Level}: scene descriptions paired with their associated utterances, where each scene represents a coherent segment reflecting an event or plot development;
    \item \textit{Role-Level}: textual profiles.
    % summarizing each role.
\end{itemize}

\paragraph{Persona Consistency.}
Existing role-playing benchmarks often rely on LLM-generated dialogue \citep{mmrole, rolellm}, which typically requires human verification or constraints to preserve personality consistency, factual grounding, and other attributes. 
In contrast, ActorMindBench is constructed from the widely known \textit{Friends} Season~1 {\footnote{\url{https://en.wikipedia.org/wiki/Friends_season_1}}, ensuring naturally consistent personas, stable character knowledge, and high-quality human-written dialogue.

\subsection{Construction Pipeline}

As illustrated in Figure \ref{fig:dataset}, the overall construction pipeline consists of three stages:

\paragraph{Utterance-Level.}
We process original audio episodes through speech denoising, diarization, and recognition to obtain clean speech segments with speaker labels and text content.
(1) Speech Denoising removes background noise, music, and environmental sounds from speech signal. 
After denoising, we obtain a clean and high-quality speech signal. 
We use {resemble-enhance}\footnote{\url{https://github.com/resemble-ai/resemble-enhance}}.
(2) Speech Diarization is the process of partitioning an speech signal containing human speech into segments based on the identity of each speaker. 
After diarization, the denoised speech signal from an entire episode can be labeled with who spoke at which time, allowing us to obtain speech segments with role labels. 
We use {pyannote-audio}\footnote{\url{https://github.com/pyannote/pyannote-audio}}.
(3) Speech Recognition converts speech into text, after which, we can extract the textual content from the speech. 
We use {Whisper} \footnote{\url{https://github.com/openai/whisper}} \citet{radford2023robust}.
After these processes, we obtain utterance-level content, including speech segments with corresponding role labels and textual content.

\paragraph{Scene-Level.}
Scene boundaries, indicating which utterance starts and ends a scene, are obtained by crawling online scripts with scene boundaries and then aligning them with the utterance context. 
Once the boundaries are identified, we use {Llama3}\footnote{\url{https://huggingface.co/meta-llama/Meta-Llama-3-8B}}\citet{llama3} to generate descriptive scene captions based on the dialogue within the scene. 
An illustration of the prompt can be seen in Figure \ref{fig:scene_gen} of Appendix \ref{appendix:AMB_prompt}.

\paragraph{Role-Level.}
To ensure the high authenticity of ActorMindBench, the roles included are well-known characters: Rachel, Monica, Phoebe, Joey, Chandler, and Ross. {Wikipedia} \footnote{\url{https://en.wikipedia.org/wiki/Main_Page}} already provides a detailed illustration of these roles. 
As shown in Figure \ref{fig:role_gen} of Appendix \ref{appendix:AMB_prompt}, we used Llama3 \citet{llama3} to summarize the Wikipedia pages to obtain the role profiles.

\subsection{Statistic}

ActorMindBench is derived from Season~1 of \textit{Friends}, which contains 24 episodes. After processing, we obtain:
\begin{itemize}[noitemsep,nolistsep,leftmargin=1em]
    \item \textit{Utterance-Level}: 7,653 utterances, corresponding to 5 hours and 15 minutes of speech;
    \item \textit{Scene-Level}: 313 scenes, with an average of 28.7 utterances and 4.23 roles per scene;
    \item \textit{Role-Level}: textual profiles for the 6 main characters.
\end{itemize}
Comprehensive episode- and role-level statistics are provided in Appendix~\ref{appendix:AMB_statistic}.

\section{ActorMind}
\label{sec:actormind}

\begin{figure*}[t!]
  \centering
  \includegraphics[width=\linewidth]{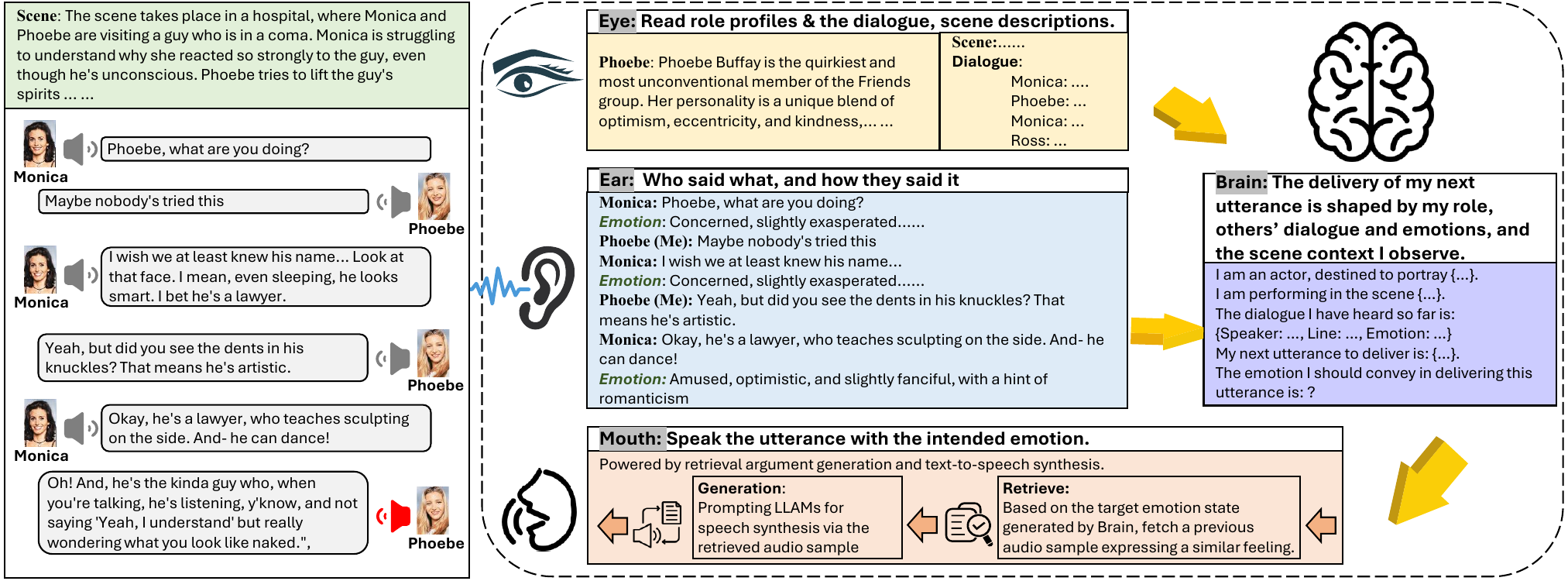}
  \caption{
  Overview of ActorMind.
  ActorMind operates in a multi-agent chain-of-thought reasoning style. 
  Specifically: 
  (1) The \textbf{\textcolor{olive}{Eye Agent}} grasps the scene descriptions and role profiles.
  (2) The \textbf{\textcolor{cyan}{Ear Agent}} listens to the tones expressed by others.
  (3) The \textbf{\textcolor{violet}{Brain Agent}} brainstorms the emotional state for the next line of dialogue based on what has been seen and heard.
  (4) Powered by RAG, the \textbf{\textcolor{orange}{Mouth Agent}} retrieves the most similar speech from its won database with a comparable emotional description, mimics it, and spontaneously delivers it.
  }
  \label{fig:AM_overview}
\end{figure*}

\subsection{Preliminaries}

\subsubsection{Notation}

\paragraph{Utterance-Level.}
We represent the utterance as 
\begin{equation}
    U = \{(U_i^{r}, U_i^{s}, U_i^{t})\}_{i=1}^{N_u},
\end{equation}
where $N_u$ denotes total utterance number, and each utterance $U_i$ is a triplet of the role indicator $U_i^{r}$, speech signal $U_i^{s}$, and the corresponding textual content $U_i^{t}$.

\paragraph{Scene-Level.}
Scene is represented as 
\begin{equation}
    S = \{(S_j^{desc}, S_j^{bd})\}_{j=1}^{N_s},
\end{equation}
where $N_s$ denotes total scene number, and each scene $S_j$ is a tuple of textual scene description $S_j^{desc}$ and scene boundary $S_j^{bd}$, which indicates the start and end points of utterances.

\paragraph{Role-Level.}
Role information is represented as 
\begin{equation}
    R = \{R_k\}_{k=1}^{N_r},
\end{equation}
where $N_r$ is total role number and $R_k$ is a textual role profile.

\subsubsection{Problem Definition}

Given scene description $S_j^{desc}$ and dialogue sequence $(U_p, \cdots, U_{q-1})$, the $U_q^{r}$-playing model should spontaneously perform the next line $\tilde{U_q^{t}}$, corresponding to the text $U_q^{t}$, in an oral manner.

\subsection{Overview}

ActorMind is a multi-agent CoT reasoning framework that facilitates speech role-playing by emulating how human actors perform in theater.  
Specifically, to conduct the role-playing of $R_k$ in scene $S_j$:  
First, the \textbf{\textcolor{olive}{Eye Agent}} grasps the scene descriptions and role profiles.
Then, the \textbf{\textcolor{cyan}{Ear Agent}} listens to the tones and emotion expressed by others.
Next, the \textbf{\textcolor{violet}{Brain Agent}} brainstorms the emotional state for the next line of dialogue based on what has been seen and heard.
Finally, powered by RAG, the \textbf{\textcolor{orange}{Mouth Agent}} retrieves a most similar speech from its own database with a comparable emotional description, mimics it, and spontaneously delivers the target line.

\subsection{Eye Agent}

In the preparatory stage, the \textbf{\textcolor{olive}{Eye Agent}} reads, context textual dialogue $(U_p^t, \cdots, U_{q-1}^t)$, the preparatory descriptive content, including the textual role profile $R_k$ and the scene description $S_j^{\text{desc}}$, and retains them in memory.

\subsection{Ear Agent}

During role-playing, empowered by Speech Emotion Captioning tools (SECAP) \citep{secap}, the \textbf{\textcolor{cyan}{Ear Agent}} listens to the dialogue sequence $(U_p^s, \cdots, U_{q-1}^s)$ and extracts the corresponding speech tone and emotional description $(E_p, \cdots, E_{q-1})$, which will be logged in textual format:  
\begin{equation}
    \begin{aligned}
    (E_p, \cdots, & E_{q-1}) = Ear[(U_p^s, \cdots, U_{q-1}^s)] \\
    &= SECAP[(U_p^s, \cdots, U_{q-1}^s)].
    \end{aligned}
\end{equation}

\subsection{Brain Agent}

The \textbf{\textcolor{violet}{Brain Agent} } serves as a central component in speech role-playing, responsible for role injection and deep contextual understanding. 
Leveraging the powerful reasoning capabilities of LLMs \citep{rolellm, llama3}, the \textbf{\textcolor{violet}{Brain Agent}} infers a reasonable emotional state $\widetilde{E_{q}}$ for delivering the next line $U_q^t$ based on what ActorMind has just perceived (seen and heard):

\begin{equation}
    \begin{aligned}
    \widetilde{E_{q}} &= Brain[R_k, S_j^{\text{desc}}, (U_p^t, E_p), \cdots, \\
                                                                            & \hspace{2cm} (U_{q-1}^t, E_{q-1}), U_q^t] \\
                    &= LLM[{Prompt}^{ear}, R_k, S_j^{\text{desc}}, (U_p^t, E_p), \cdots, \\
                                                                            & \hspace{2cm} (U_{q-1}^t, E_{q-1}), U_q^t]. \\
    \end{aligned}
\end{equation}

\subsection{Mouth Agent}

Supported by RAG, the \textbf{\textcolor{orange}{Mouth Agent}} retrieves a previously performed speech segment $U_x^s$ from its database $Database_{U_k}$ whose emotional state is most similar to $\widetilde{E_{q}}$.  
Leveraging the in-context learning capability of TTS models, the agent is then prompted with the target text $U_q^t$ together with the retrieved speech $U_x^s$, enabling it to render the target utterance with the voice and emotional tone of the retrieved sample:
\begin{equation}
    \begin{aligned}
    \widetilde{U_q^{s}} &= Mouth(\widetilde{E_{q}}, Database_{U_k}, U_q^t)\\
                        &=RAG(\widetilde{E_{q}}, Database_{U_k}, U_q^t). \\
    \end{aligned}
\end{equation}

\section{Experiment}

\subsection{Dataset}

ActorMindBench is constructed from \textit{Friends Season 1} (24 episodes), with details on structure and statistics illustrated in Section \ref{sec:actormindbench}. 
Episodes 1–10 and 15–24 are used for training and deployment, while episodes 11–14 are reserved for testing. 
This split ensures that the training and deployment data encompass a broad range of emotional expressions, from relatively neutral states in earlier episodes to higher-intensity emotions in later episodes, thereby supporting robust role modeling.

\subsection{Evaluation Metric}
We utilize the mean opinion score (MOS) \citep{chu2006objective} to measure the perceived quality of the generated speech. 
To adapt this metric for the role-playing setting, we introduce the RP-MOS. 
It ranges from 1 to 5, with 1 indicating the lowest quality and 5 the highest. 
In the speech role-playing context, we identify two pivotal aspects: (1) \textbf{Exact Delivery} and (2) \textbf{Emotion Expression}.

\paragraph{Exact Delivery.} It refers to the accurate impersonation of the intended character's voice and the precise articulation of target words. This aspect is fundamental to role-playing; without it, effective speech role-playing may not be feasible. 
Given this landscape, we consider Exact Delivery a prerequisite capability. In our RP-MOS, if the generated speech fails to resemble the intended character's voice or does not convey the correct content, we assign a lowest score of 1.

\paragraph{Emotion Expression.}
As the adage goes, “there are a thousand Hamlets in a thousand people’s eyes”—human interpretation of emotion expression can vary greatly. 
For research purposes, however, a clear and reproducible evaluation criterion is crucial. 
Thus, we consider the emotional expression evident in the original speech segments—reflected through prosodic cues such as tone, tempo, and intensity—as the ground truth proxy for role-playing emotion expression quality. 
By evaluating the emotional alignment between the model's output and this ground truth, we can capture a measurable dimension of emotion expression: the model's ability to reproduce the intended expressive stance of a character within a specific scene. 

Further details on the RP-MOS instructions and evaluator guidelines are provided in Appendix \ref{appendix:EXP_metric}.

\subsection{Baselines}
\label{appendix:Exp_baselines}

We evaluate \textbf{ActorMind} against six baseline methods. These methods are grouped into two categories: (1) LLAM, and (2)–(6) TTS models.
Specifically, 
(1). \textbf{Qwen\_Omni} \citep{qwenomni} is a multimodal foundation model capable of processing and generating text, images, and audio, supporting real-time, multilingual, and multimodal interactions
We report results using the official 7B checkpoint.\footnote{\url{https://huggingface.co/Qwen/Qwen2.5-Omni-7B}}.
The prompt used for Qwen\_Omni speech role-playing is shown in Figure~\ref{fig:bsl_qwen_prompt} (Appendix~\ref{appedex:qwen}).
(2). \textbf{CosyVoice} \citep{cosyvoice}  is a semantic-codec-based TTS model. 
We evaluate the official 0.5B checkpoint.\footnote{\url{https://www.modelscope.cn/studios/qaz321456/CosyVoice2-0.5B}}
(3). \textbf{SparkTTS} \citep{sparktts}  is an efficient TTS model that combines a single-stream disentangled codec with an LLM backbone. 
We evaluate the official 0.5B checkpoint.\footnote{\url{https://huggingface.co/SparkAudio/Spark-TTS-0.5B}}
(4). \textbf{IndexTTS} \citep{indextts} is an autoregressive zero-shot TTS model with controllable duration and expressive emotion modeling. 
We evaluate the official $\sim$0.5B checkpoint.\footnote{\url{https://huggingface.co/IndexTeam/IndexTTS-2}}
(5). \textbf{YourTTS} \citep{yourtts} is a flow-matching-based generative model for high-quality speech synthesis. 
We evaluate the official $\sim$90M checkpoint.\footnote{\url{https://github.com/Edresson/YourTTS}}
(6). \textbf{F5-TTS} \citep{f5-tts} is a fully non-autoregressive TTS model based on flow matching with a Diffusion Transformer. 
We evaluate the official 300M checkpoint.\footnote{\url{https://huggingface.co/SWivid/F5-TTS/tree/main/F5TTS_Base}}

\subsection{Implementation}

Please refer to  Appendix \ref{appendix:AM_tool}.

\section{Results and Analysis}

\subsection{Main Result}

Subjective evaluation using RP-MOS is presented in Tables \ref{tab:Result_main_mos}.

\begin{table*}[htbp]
  \centering
  \resizebox{\linewidth}{!}{
  \begin{tabular}{l|cccccc|c}
    \hline
         & Phoebe & Joey & Chandler & Rachel & Ross & Monica & Average \\ 
         \hline
         % \cline{2-8}
        YourTTS \citep{yourtts} & 2.90 ± 0.89 & 2.47 ± 0.90 & 2.30 ± 1.40 & 1.80 ± 0.84 & 2.60 ± 1.19 & 2.30 ± 0.91 & 2.39 ± 0.93 \\
        F5-TTS \citep{f5-tts} & 2.60 ± 1.08 & 2.33 ± 0.75 & \textbf{3.60 ± 0.65} & 3.00 ± 1.58 & 2.90 ± 0.74 & 2.80 ± 0.45 & 2.87 ± 0.77 \\ 
        Cosyvoice \citep{cosyvoice} & 2.30 ± 0.76 & 2.67 ± 0.71 & 2.10 ± 0.22 & 1.40 ± 0.55 & 2.00 ± 0.94 & 1.80 ± 0.67 & 2.04 ± 0.45 \\ 
        SparkTTS \citep{sparktts} & 3.40 ± 1.08 & 2.53 ± 0.80 & 2.90 ± 0.42 & 2.20 ± 1.30 & 3.20 ± 0.91 & 2.00 ± 0.94 & 2.71 ± 0.78 \\ 
        IndexTTS \citep{indextts} & 3.80 ± 0.67 & 2.20 ± 0.65 & 3.30 ± 0.27 & 3.20 ± 0.84 & 2.60 ± 1.08 & 3.20 ± 0.76 & 3.05 ± 0.56 \\ 
        Qwen\_Omni \citep{qwenomni} & 1.00 ± 0.00 & 1.00 ± 0.00 & 1.00 ± 0.00 & 1.00 ± 0.00 & 1.00 ± 0.00 & 1.00 ± 0.00 & 1.00 ± 0.00 \\ 
        \hdashline
        ActorMind (Ours) & \textbf{4.00 ± 0.05} & \textbf{3.47 ± 0.61} & 3.20 ± 0.45 & \textbf{3.40 ± 0.89} & \textbf{3.70 ± 0.57} & \textbf{3.60 ± 0.55} & \textbf{3.56 ± 0.27}  \\ \hline
  \end{tabular}
  }
  
  \caption{
    \textbf{Main Results.} 
    Subjective evaluation using RP-MOS for ActorMind and baseline models. 
  }
  \label{tab:Result_main_mos}
\end{table*}

\begin{itemize}[noitemsep,nolistsep,leftmargin=1em]
    \item Overall, on the average score across all roles, ActorMind demonstrates superior performance, outperforming all baseline LLAMs and TTS models. 
    This indicates that, in speech role-playing scenarios, ActorMind effectively considers role profiles, scenes, and dialogue to respond spontaneously—capabilities not present in current models. 
    This positions ActorMind as a pioneering model in the realm of speech role-playing, advancing the field significantly.
    \item Among the roles, ActorMind does not achieve optimal performance for Chandler in the subjective evaluation.
    % , nor for Phoebe, Chandler and Joey in the objective evaluation. 
    This may be attributed to their vivid and diverse speaking styles, which demand more advanced reasoning capabilities and meticulous design in future models.
    \item Among all models evaluated, Qwen\_Omni exhibited the poorest performance. 
    There are several contributing factors: 
    (1) The limited set of voices provided by Qwen\_Omni does not align with the roles on ActorMindBench;
    % , which are derived from Friends; 
    (2) Qwen\_Omni is primarily designed for multimodal understanding, resulting in many generated speech segments that are neutral and lacking in expressiveness necessary for role-playing; 
    and (3) During our experiments, when prompts were long—incorporating role profiles and contextual details—Qwen\_Omni struggled to accurately express the intended content, rendering it unsuitable for role-playing applications.
\end{itemize}

\begin{table*}[t]
  \centering
  \resizebox{\linewidth}{!}{
  \begin{tabular}{l|cccccc|c}
    \hline
         & Phoebe & Joey & Chandler & Rachel & Ross & Monica & ALL \\ 
         \hline
         % \cline{2-8}
       ActorMind + F5-TTS  & \textbf{1.00 ± 0.00} & 0.75 ± 0.29 & 0.75 ± 0.50 & 0.50 ± 0.58 & 0.88 ± 0.25 & 0.75 ± 0.29 & 0.77 ± 0.18 \\ 
       ActorMind + Cosyvoice  & 0.88 ± 0.25 & 0.63 ± 0.25 & 0.75 ± 0.29 & 0.50 ± 0.58 & \textbf{\textit{0.38 ± 0.48}} & 0.63 ± 0.48 & 0.63 ± 0.16 \\ 
       ActorMind + SparkTTS  & 0.50 ± 0.00 & 0.88 ± 0.25 & \textbf{1.00 ± 0.00} & \textbf{1.00 ± 0.00} & \textbf{1.00 ± 0.00} & \textbf{1.00 ± 0.00} & 0.90 ± 0.04 \\ 
       ActorMind + IndexTTS  & 0.88 ± 0.25 & 0.75 ± 0.5 & \textbf{\textit{0.25 ± 0.50}} & 0.75 ± 0.50 & 0.88 ± 0.25 & \textbf{1.00 ± 0.00} & 0.75 ± 0.17 \\ 
       ActorMind + YourTTS  & 0.63 ± 0.48 & 0.50 ± 0.41 & 0.88 ± 0.25 & 0.50 ± 0.58 & 1.00 ± 0.00 & 0.50 ± 0.58 & 0.67 ± 0.17 \\ 
        \hline
  \end{tabular}
  }
  
  \caption{
    Performance improvement over baseline models after applying ActorMind.
    }
  \label{tab:Result_ablation}
\end{table*}

\subsection{Ablation Study}

\begin{table}[htbp]
  \centering
  \resizebox{.9\linewidth}{!}{
  \begin{tabular}{llcc}
    \hline
           & & RP-MOS $\uparrow$        \\ \hline
        % ActorMind                               & 0                  \\ 
        (1) &    w/o Role Profile (w/o Eye)              & -0.37 ± 0.21      \\ 
        (2) &    w/o Scene (w/o Eye)                     & -0.30 ± 0.17     \\ 
        (3) &    w/o Context (w/o Eye, w/o Ear)          & -0.22 ± 0.14      \\ 
        (4) &    w/o Ear                                 & -0.32 ± 0.23      \\ 
        (5) &    w/o Brain (w/o All)                     & -0.51 ± 0.56      \\ 
        \hline
  \end{tabular}
  }
  \caption{
    \textbf{Ablation study.} Relative performance with respect to ActorMind across six roles.
    }
  \label{tab:ablation}
\end{table}

We conduct ablation studies to evaluate the contribution of each agent in ActorMind, as well as the necessity of the role profile, scene description, and context in the speech role-playing setting.
ActorMind operates as a sequential pipeline.
Therefore, removing any component may disrupt this pipeline, leading to interdependent effects in the ablation study. For example, removing the Brain agent effectively disables the RAG mechanism in the Mouth agent.

The Eye agent provides all essential textual inputs for role-playing, including the role profile, scene description, and context textual lines. 
To assess its contribution and the necessity of its inputs, we conduct three ablation settings: (1) \textbf{w/o Role Profile (w/o Eye)}, (2). \textbf{w/o Scene (w/o Eye)}, (3). \textbf{w/o Context (w/o Eye, w/o Ear)}.
In (3), removing the textual context eliminates dialogue-based speech emotion processing; therefore, the Ear agent is also removed.

The Ear agent processes speech emotion information. 
To evaluate its effect, we conduct:
(4). \textbf{w/o Ear}, where only textual context is used without speech emotion cues.

The Brain agent infers the emotion of the target utterance. 
Without the Brain agent, RAG in the Mouth agent— which relies on inferred emotions to retrieve speech prompts for TTS—cannot function. 
Moreover, information from the Eye and Ear agents becomes ineffective. 
Therefore:
(5). \textbf{w/o Brain (w/o All)}, which corresponds to \textbf{w/o Eye, w/o Ear, w/o Brain, w/o Mouth}.

We report relative performance with respect to ActorMind across six roles. 
Subjective evaluation is measured by the average RP-MOS difference.

As shown in Table \ref{tab:ablation}, the results of (1)-(3) indicate that removing any of these components leads to performance degradation, highlighting their importance in speech role-playing setting. 
Among them, removing the role profile results in the largest performance drop, demonstrating that role profile information is the most critical component for speech role-playing. 
This observation aligns with intuitive expectations for role-conditioned generation.

Overall, the results from settings (1)–(5) show that each component in ActorMind is necessary, validating the soundness of our method design.

\subsection{Generalization of ActorMind}

ActorMind is a multi-agent CoT reasoning framework. 
It operates in an off-the-shelf manner without requiring additional training, making generalization a core consideration. 
To evaluate this property, we replace the speech generation component with different models, and compare ActorMind + [MODEL] against each corresponding standalone model, thereby assessing ActorMind’s effectiveness as a universal reasoning framework. 
We intentionally omit Qwen\_Omni, as it does not support target voice generation and therefore cannot support role-playing.

In this experiment, we conduct subjective evaluations, where evaluators assign a score of 1 to indicate a clear improvement, 0.5 to indicate equivalence, and 0 to indicate degradation relative to the baseline model. We recruit six English-speaking evaluators for this study.

As shown in Table \ref{tab:Result_ablation}, except for ActorMind + CosyVoice on Ross and ActorMind + IndexTTS on Chandler, all ActorMind + [MODEL] configurations achieve scores higher than 0.5, demonstrating consistent performance gains over their corresponding baselines. Moreover, five configurations achieve a score of 1, indicating absolute improvement and further highlighting the effectiveness and robustness of ActorMind as a general-purpose reasoning method.

\subsection{Qualitative Analysis}

We visualize spectrograms of the generated speech to qualitatively evaluate speech role-playing ability. In each spectrogram, the x-axis represents time, reflecting the temporal dynamics and tone, while the y-axis represents the energy distribution across frequency bins, reflecting the vocal characteristics. Higher similarity to the ground-truth spectrogram indicates better alignment in both prosody and speaker-specific traits.

\begin{figure}[t]
    \centering

    % Row 1
    \begin{subfigure}[t]{0.32\linewidth}
        \centering
        \includegraphics[width=\linewidth]{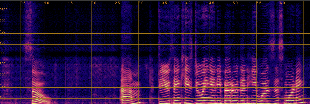}
        \caption{CosyVoice}
    \end{subfigure}
    \hfill
    \begin{subfigure}[t]{0.32\linewidth}
        \centering
        \includegraphics[width=\linewidth]{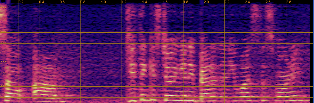}
        \caption{SparkTTS}
    \end{subfigure}
    \hfill
    \begin{subfigure}[t]{0.32\linewidth}
        \centering
        \includegraphics[width=\linewidth]{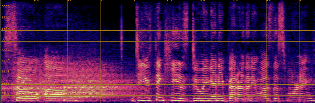}
        \caption{IndexTTS}
    \end{subfigure}

    \vspace{0.4em}

    % Row 2

    \begin{subfigure}[t]{0.32\linewidth}
        \centering
        \includegraphics[width=\linewidth]{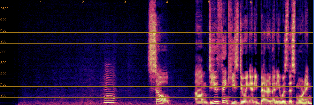}
        \caption{F5-TTS}
    \end{subfigure}
    \hfill
    \begin{subfigure}[t]{0.32\linewidth}
        \centering
        \includegraphics[height=0.82cm, width=\linewidth]{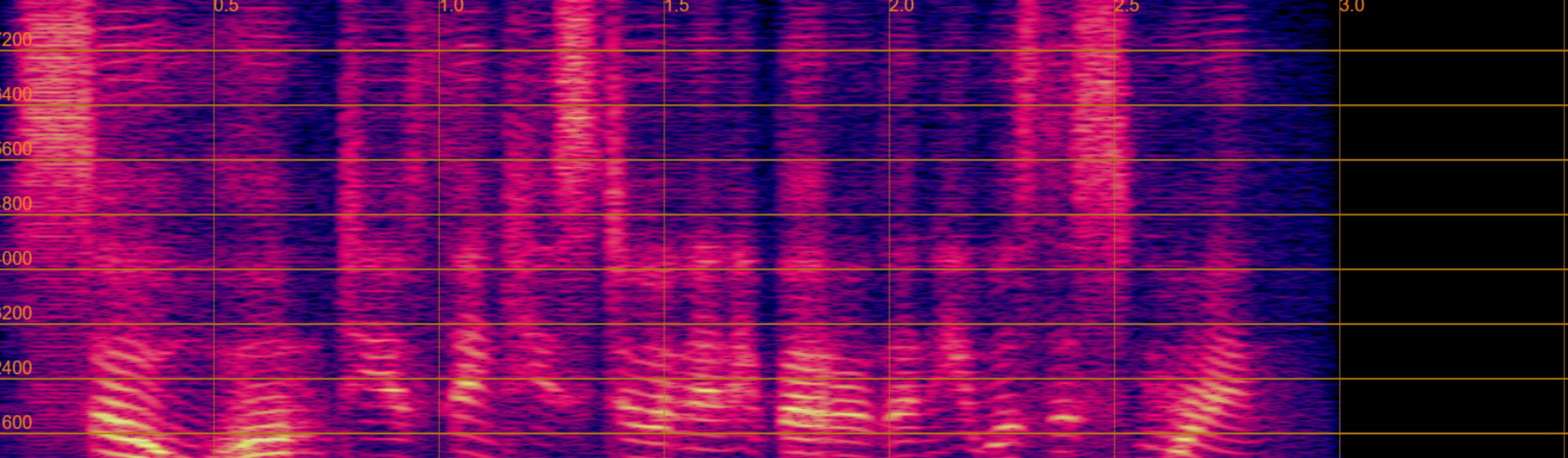}
        \caption{YourTTS}
    \end{subfigure}
    \hfill
    \begin{subfigure}[t]{0.32\linewidth}
        \centering
        \includegraphics[width=\linewidth]{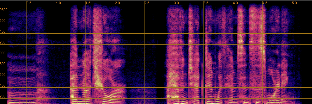}
        \caption{Qwen-Omni}
    \end{subfigure}

    \vspace{0.4em}
    \begin{subfigure}[t]{0.32\linewidth}
        \centering
        \includegraphics[width=\linewidth]{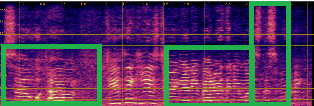}
        \caption{ActorMind}
    \end{subfigure}
    % \hfill
    \hspace{0.01\linewidth}
    \begin{subfigure}[t]{0.32\linewidth}
        \centering
        \includegraphics[width=\linewidth]{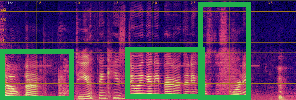}
        \caption{Ground Truth}
    \end{subfigure}

    \caption{
    Spectrogram Comparison of baselines and ActorMind.
    All samples are generated for \textit{Phoebe} performing \textit{"...So, um, do you think he's doing any better than he was this morning?"} under the same scene and context.
    }
    \label{fig:Result_case}
\end{figure}

Figure \ref{fig:Result_case} presents spectrograms of the generated outputs alongside the corresponding ground-truth speech. As shown:
\begin{itemize}[noitemsep,nolistsep,leftmargin=1em]
    \item TTS models (Figure \ref {fig:Result_case} (a)–(e)) use randomly sampled prompts, resulting in arbitrary tone and prosody. Although these models can successfully generate the target utterance with the target voice, their energy distributions over time and frequency differ substantially from the ground truth. In contrast, as highlighted by the green boxes, ActorMind exhibits significantly higher spectrogram similarity, indicating more accurate role-consistent prosody and expression.
    \item LLAM, i.e., Qwen\_Omni (Figure \ref {fig:Result_case} (f)), fails to reproduce the target voice. This is reflected in its energy distribution across the frequency axis, which deviates significantly from those of the other models and indicates a mismatch in speaker characteristics.
\end{itemize}

\section{Conclusion}

To establish speech role-playing, we formalize the concept of speech role-playing and introduce ActorMindBench, a public benchmark along with corresponding construction pipeline.
We also propose ActorMind, a multi-agent CoT style reasoning framework,
which is off-the-shelf and can be applied without additional training.

Experimental results on ActorMindBench demonstrate the effectiveness of ActorMind compared to baseline models. Additional experiments show that ActorMind is effective as a universal framework across different speech generation models, and qualitative analyses on spectrograms further illustrate its ability to produce spontaneous speech.

\section{Limitations}

\paragraph{ActorMindBench.}
ActorMindBench is entirely derived from Friends Season 1, covering six roles within the urban comedy domain. As such, it has a limited set of roles and domain coverage. 
Despite these limitations, ActorMindBench offers a valuable test bed for current research. 
Notably, as shown in the main results, current methods perform poorly in speech role-playing setting, indicating that, although ActorMindBench is limited, it is sufficient for researchers to explore and develop new approaches.

\paragraph{ActorMind.}
ActorMind is an off-the-shelf method that does not require any training. While it demonstrates strong performance, further improvements may be possible through further training, for example, using reinforcement learning to enhance the RAG mechanism in the Mouth agent or to improve emotion reasoning in the Brain agent. Nevertheless, as the first system of its kind, ActorMind represents a meaningful step forward in speech role-playing.

\section{Ethical Considerations}

ActorMindBench benchmark is built upon the well-known TV series \textit{Friends, Season 1}, and includes annotations at the utterance, scene, and role levels. 
However, \textit{Friends} is protected by copyright.
Accordingly, we do not—and will not—distribute any copyrighted audio content from the series.

All annotations in ActorMindBench are generated using publicly available tools and consist exclusively of structured annotations. We will publicly release only the annotation files, enabling researchers to freely access and use our annotations and independently obtain the original Friends episodes through legitimate channels for their own research purposes.
This design strictly follows established community practice and ensures that no copyrighted media is redistributed, thereby avoiding any copyright or licensing violations.

In summary, our work:
\begin{itemize}[noitemsep,nolistsep,leftmargin=1em]
    \item Does not distribute any copyrighted audio content;
    \item Releases only copyright-safe annotations;
    \item Supports reproducibility and extensibility for future research;
\end{itemize}
We therefore believe that the ethical and legal usage of the source material in our benchmark is appropriate and rigorously handled.

\bibliography{custom}

\appendix

\newpage

\section{ActorMindBench Details}
\label{appendix:ActorMindBench_detail}

\subsection{ActorMindBench Example Data}  
\label{appendix:AMB_exampledata}
Example of ActorMindBench data is shown in Figure \ref{fig:AMB_example}.

\begin{figure*}[htbp]
  \centering
  \includegraphics[width=\linewidth]{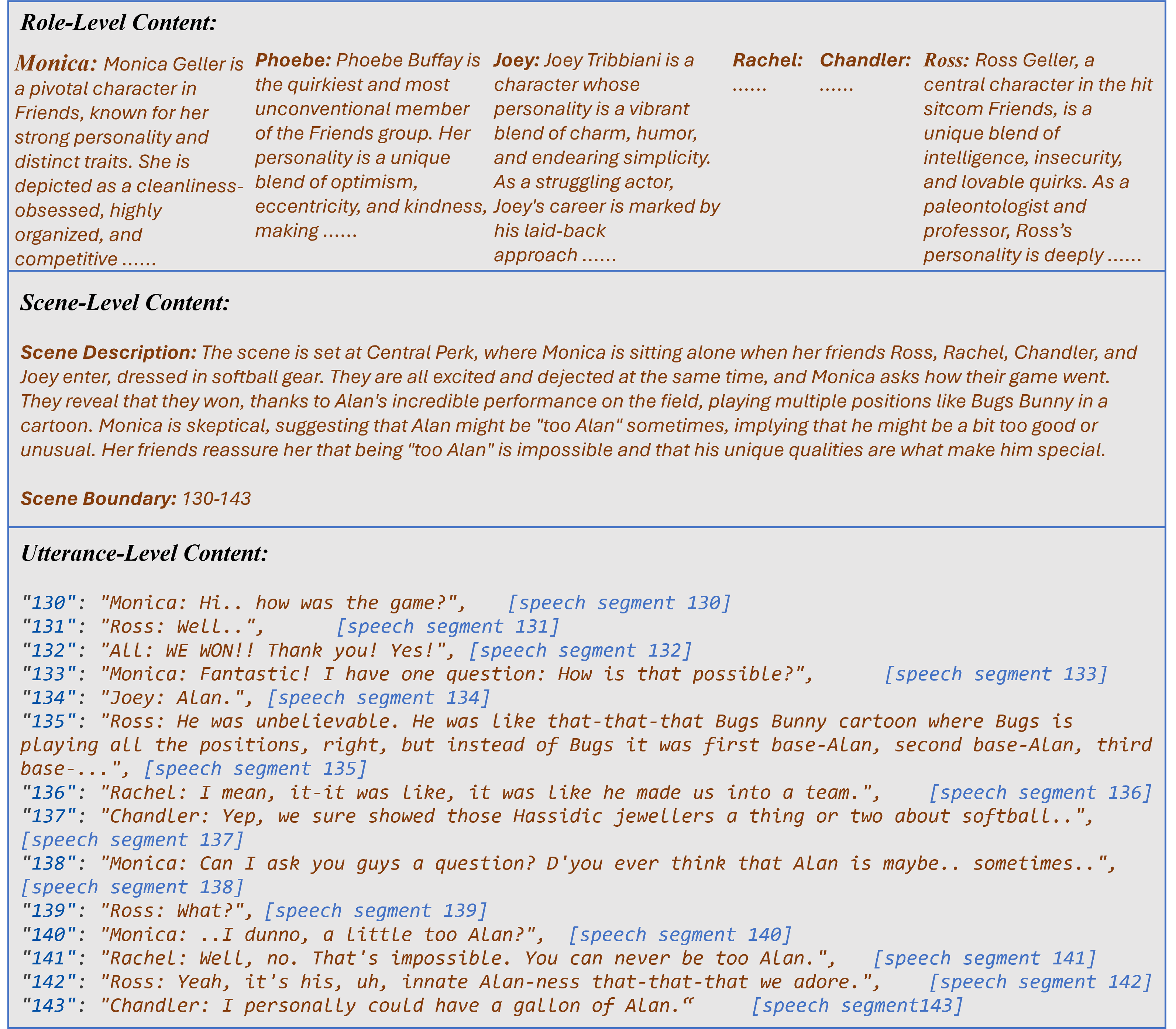}
  \caption{
  ActorMindBench Example Data
  }
  \label{fig:AMB_example}
\end{figure*}

\subsection{Prompts in ActorMindBench Construction} 
\label{appendix:AMB_prompt}

Prompts for scene summarization and role summarization can be seen in Figure \ref{fig:scene_gen} and Figure \ref{fig:role_gen} separately.

\begin{figure}[htbp]
  \centering
  \includegraphics[width=\linewidth]{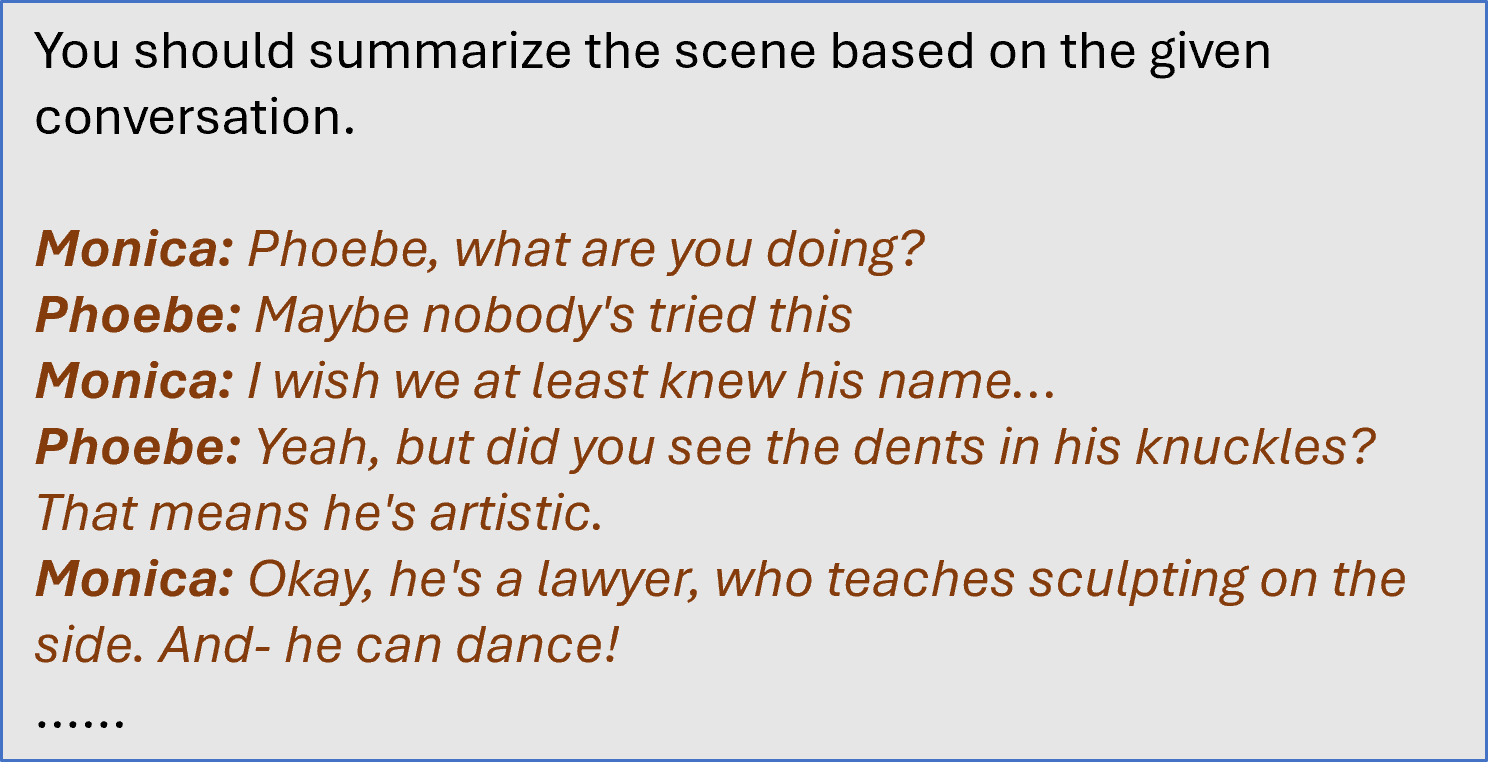}
  \caption{
  Prompt for scene captioning: black text indicates the prompt, while brownish-yellow highlights the utterances within a scene.
  }
  \label{fig:scene_gen}
\end{figure}

\begin{figure}[htbp]
  \centering
  \includegraphics[width=\linewidth]{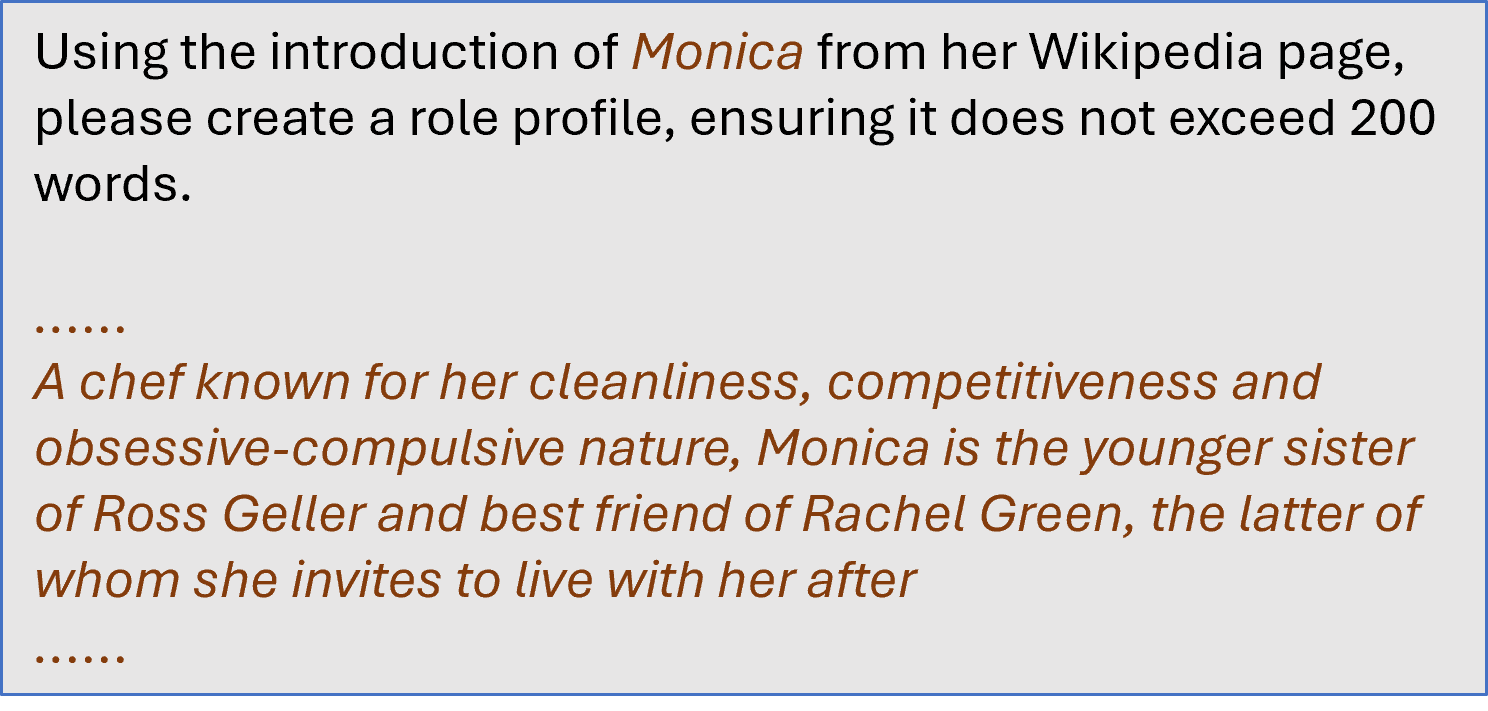}
  \caption{
  Prompt for role profile generation: black text indicates the prompt, while brownish-yellow highlights the role content.
  }
  \label{fig:role_gen}
\end{figure}

\subsection{ActorMindBench Detail Statistic}  
% Detailed Statistics of ActorMindBench
\label{appendix:AMB_statistic}

\textit{Utterance-Level} statistics regarding the number and duration of utterances for role 'xx' in episode 'yy' are provided in Table \ref{tab:AMB_utterance_sta}.

\begin{table*}[htbp]
  \centering
  \resizebox{\linewidth}{!}{
  % \begin{tabular}{l|cc|cc|cc|cc|cc|cc|cc|cc}
  \begin{tabular}{lcc|cc|cc|cc|cc|cc|cc|cc}
        \hline
         Episode & \multicolumn{2}{c}{\textbf{Rachel}} &\multicolumn{2}{c}{\textbf{Monica}}  & \multicolumn{2}{c}{\textbf{Phoebe}}  & \multicolumn{2}{c}{\textbf{Joey}} &\multicolumn{2}{c}{\textbf{Chandler}} & \multicolumn{2}{c}{\textbf{Ross}}   & \multicolumn{2}{c}{\textbf{OTHERS}} & \multicolumn{2}{c}{\textbf{TOTAL}} \\ 
         % \cline{2-17}
         \hline
         & num & duration & num & duration & num & duration & num & duration & num & duration & num & duration & num & duration & num & duration \\
         \cline{2-17}

        SE01\_01 & 86 & 0:03:15 & 82 & 0:03:08 & 22 & 0:00:54 & 39 & 0:01:42 & 33 & 0:01:23 & 63 & 0:02:45 & 25 & 0:01:07 & 350 & 0:14:13 \\ 
        SE01\_02 & 56 & 0:02:36 & 37 & 0:01:15 & 21 & 0:00:44 & 11 & 0:00:23 & 29 & 0:01:10 & 101 & 0:04:16 & 97 & 0:03:44 & 352 & 0:14:09 \\ 
        SE01\_03 & 32 & 0:01:12 & 76 & 0:02:36 & 65 & 0:02:20 & 28 & 0:01:04 & 72 & 0:02:37 & 53 & 0:01:54 & 39 & 0:01:31 & 365 & 0:13:14 \\ 
        SE01\_04 & 81 & 0:03:16 & 65 & 0:02:27 & 47 & 0:01:52 & 36 & 0:01:15 & 49 & 0:01:55 & 57 & 0:02:36 & 36 & 0:01:23 & 371 & 0:14:44 \\ 
        SE01\_05 & 48 & 0:02:01 & 40 & 0:01:44 & 29 & 0:00:57 & 58 & 0:02:23 & 50 & 0:01:51 & 73 & 0:03:30 & 48 & 0:02:05 & 346 & 0:14:31 \\ 
        SE01\_06 & 22 & 0:00:57 & 54 & 0:02:12 & 22 & 0:00:48 & 52 & 0:02:15 & 114 & 0:04:22 & 32 & 0:01:28 & 52 & 0:02:02 & 348 & 0:14:04 \\ 
        SE01\_07 & 49 & 0:02:00 & 21 & 0:00:43 & 44 & 0:01:52 & 38 & 0:01:26 & 61 & 0:02:47 & 71 & 0:03:04 & 27 & 0:00:52 & 311 & 0:12:45 \\ 
        SE01\_08 & 23 & 0:01:00 & 45 & 0:01:34 & 27 & 0:00:59 & 15 & 0:00:31 & 60 & 0:02:09 & 75 & 0:03:07 & 94 & 0:03:41 & 339 & 0:13:02 \\ 
        SE01\_09 & 64 & 0:02:28 & 74 & 0:02:45 & 31 & 0:01:13 & 43 & 0:01:28 & 54 & 0:02:07 & 65 & 0:02:43 & 34 & 0:01:20 & 365 & 0:14:04 \\ 
        SE01\_10 & 32 & 0:01:28 & 22 & 0:00:50 & 58 & 0:02:21 & 20 & 0:00:46 & 54 & 0:01:54 & 53 & 0:02:10 & 77 & 0:03:20 & 316 & 0:12:49 \\ 
        SE01\_11 & 28 & 0:01:03 & 48 & 0:01:46 & 49 & 0:01:54 & 44 & 0:01:41 & 53 & 0:01:46 & 78 & 0:02:43 & 66 & 0:02:23 & 366 & 0:13:15 \\ 
        SE01\_12 & 49 & 0:02:04 & 38 & 0:01:29 & 49 & 0:01:51 & 36 & 0:01:26 & 39 & 0:01:22 & 71 & 0:02:45 & 30 & 0:01:18 & 312 & 0:12:14 \\ 
        SE01\_13 & 26 & 0:01:15 & 15 & 0:00:30 & 27 & 0:01:16 & 52 & 0:02:12 & 41 & 0:01:35 & 14 & 0:00:41 & 106 & 0:05:01 & 281 & 0:12:30 \\ 
        SE01\_14 & 19 & 0:00:52 & 20 & 0:00:50 & 17 & 0:00:50 & 32 & 0:01:14 & 51 & 0:01:55 & 53 & 0:02:31 & 83 & 0:03:49 & 275 & 0:12:02 \\ 
        SE01\_15 & 25 & 0:00:57 & 44 & 0:02:04 & 39 & 0:01:34 & 32 & 0:01:15 & 70 & 0:02:55 & 35 & 0:01:32 & 24 & 0:01:01 & 269 & 0:11:19 \\ 
        SE01\_16 & 27 & 0:01:08 & 13 & 0:00:36 & 41 & 0:02:03 & 22 & 0:01:01 & 59 & 0:02:34 & 36 & 0:01:39 & 75 & 0:03:17 & 273 & 0:12:17 \\ 
        SE01\_17 & 54 & 0:02:05 & 63 & 0:02:40 & 33 & 0:01:29 & 30 & 0:01:13 & 25 & 0:01:04 & 50 & 0:02:12 & 102 & 0:03:56 & 357 & 0:14:38 \\ 
        SE01\_18 & 84 & 0:03:42 & 38 & 0:01:31 & 34 & 0:01:25 & 21 & 0:00:55 & 33 & 0:01:16 & 59 & 0:02:18 & 9 & 0:00:20 & 278 & 0:11:27 \\ 
        SE01\_19 & 91 & 0:04:04 & 35 & 0:01:21 & 18 & 0:00:45 & 19 & 0:00:51 & 27 & 0:01:06 & 88 & 0:03:58 & 40 & 0:01:38 & 318 & 0:13:43 \\ 
        SE01\_20 & 85 & 0:03:51 & 30 & 0:01:13 & 21 & 0:00:48 & 34 & 0:01:32 & 62 & 0:02:32 & 22 & 0:01:05 & 69 & 0:02:57 & 323 & 0:13:58 \\ 
        SE01\_21 & 34 & 0:01:26 & 64 & 0:03:03 & 11 & 0:00:34 & 21 & 0:00:57 & 25 & 0:01:10 & 51 & 0:02:35 & 67 & 0:03:09 & 273 & 0:12:53 \\ 
        SE01\_22 & 27 & 0:01:05 & 50 & 0:02:16 & 53 & 0:02:07 & 16 & 0:00:38 & 50 & 0:02:03 & 41 & 0:01:42 & 40 & 0:01:53 & 277 & 0:11:44 \\ 
        SE01\_23 & 24 & 0:01:02 & 25 & 0:00:55 & 35 & 0:01:40 & 34 & 0:01:28 & 22 & 0:00:57 & 68 & 0:02:44 & 102 & 0:04:23 & 310 & 0:13:09 \\ 
        SE01\_24 & 64 & 0:02:59 & 35 & 0:01:39 & 19 & 0:00:54 & 62 & 0:02:30 & 28 & 0:01:15 & 36 & 0:01:34 & 34 & 0:01:38 & 278 & 0:12:28 \\ \hdashline
        \textbf{ALL} & \textbf{1130} & \textbf{0:47:47} & \textbf{1034} & \textbf{0:41:06} & \textbf{812} & \textbf{0:33:09} & \textbf{795} & \textbf{0:32:06} & \textbf{1161} & \textbf{0:45:43} & \textbf{1345} & \textbf{0:57:31} & \textbf{1376} & \textbf{0:57:50} & \textbf{7,653} & \textbf{5:15:12} \\ 
        \hline
    \end{tabular}
  }
  \caption{
    Utterance-level statistics on number of utterances and speech duration by role and episode.
  }
  \label{tab:AMB_utterance_sta}
\end{table*}

\textit{Scnene-Level} statistics regarding the average number of utterances and roles performed per scene in each episode are provided in Table \ref{tab:AMB_scene_sta}.

\begin{table}[htbp]
  \centering
  \resizebox{\linewidth}{!}{
  \begin{tabular}{lccc}
    \hline
        Episode & \textbf{Scene Num} & \textbf{Avg Utterances per Scene} & \textbf{Avg Roles per Scene} \\ 
        \hline
        SE01\_01 & 14 & 17.29 & 3.93 \\ 
        SE01\_02 & 8 & 29.75 & 5.38 \\ 
        SE01\_03 & 13 & 20.0 & 4.85 \\ 
        SE01\_04 & 16 & 15.75 & 4.19 \\ 
        SE01\_05 & 16 & 14.94 & 3.31 \\ 
        SE01\_06 & 9 & 24.33 & 4.78 \\
        SE01\_07 & 21 & 11.14 & 2.95 \\ 
        SE01\_08 & 10 & 16.9 & 4.50 \\ 
        SE01\_09 & 12 & 19.08 & 3.92 \\ 
        SE01\_10 & 8 & 29.0 & 6.00 \\
        SE01\_11 & 12 & 23.92 & 4.50 \\ 
        SE01\_12 & 15 & 17.33 & 4.33 \\ 
        SE01\_13 & 13 & 18.69 & 4.31 \\
        SE01\_14 & 17 & 11.12 & 3.41 \\ 
        SE01\_15 & 14 & 17.43 & 3.43 \\ 
        SE01\_16 & 14 & 19.5 & 5.07 \\ 
        SE01\_17 & 14 & 20.14 & 4.14 \\ 
        SE01\_18 & 8 & 33.38 & 6.25 \\ 
        SE01\_19 & 8 & 31.38 & 5.12 \\ 
        SE01\_20 & 12 & 20.33 & 4.92 \\ 
        SE01\_21 & 15 & 14.07 & 4.00 \\ 
        SE01\_22 & 12 & 21.42 & 4.00 \\ 
        SE01\_23 & 21 & 12.76 & 4.1 \\ 
        SE01\_24 & 11 & 23.91 & 4.00 \\ 
         \hdashline
        ALL & \textbf{313} & \textbf{18.7} & \textbf{4.23} \\ \hline
    \end{tabular}
  }
  \caption{
    Scene-level statistics.
  }
  \label{tab:AMB_scene_sta}
\end{table}

\section{Experiment Details}
\label{exp}

\subsection{Qwen\_Omni in Speech Role-Playing}  
\label{appedex:qwen}

The prompt used for Qwen\_Omni speech role-playing is shown in Figure~\ref{fig:bsl_qwen_prompt}.

\begin{figure}[htbp]
  \centering
  \includegraphics[width=\linewidth]{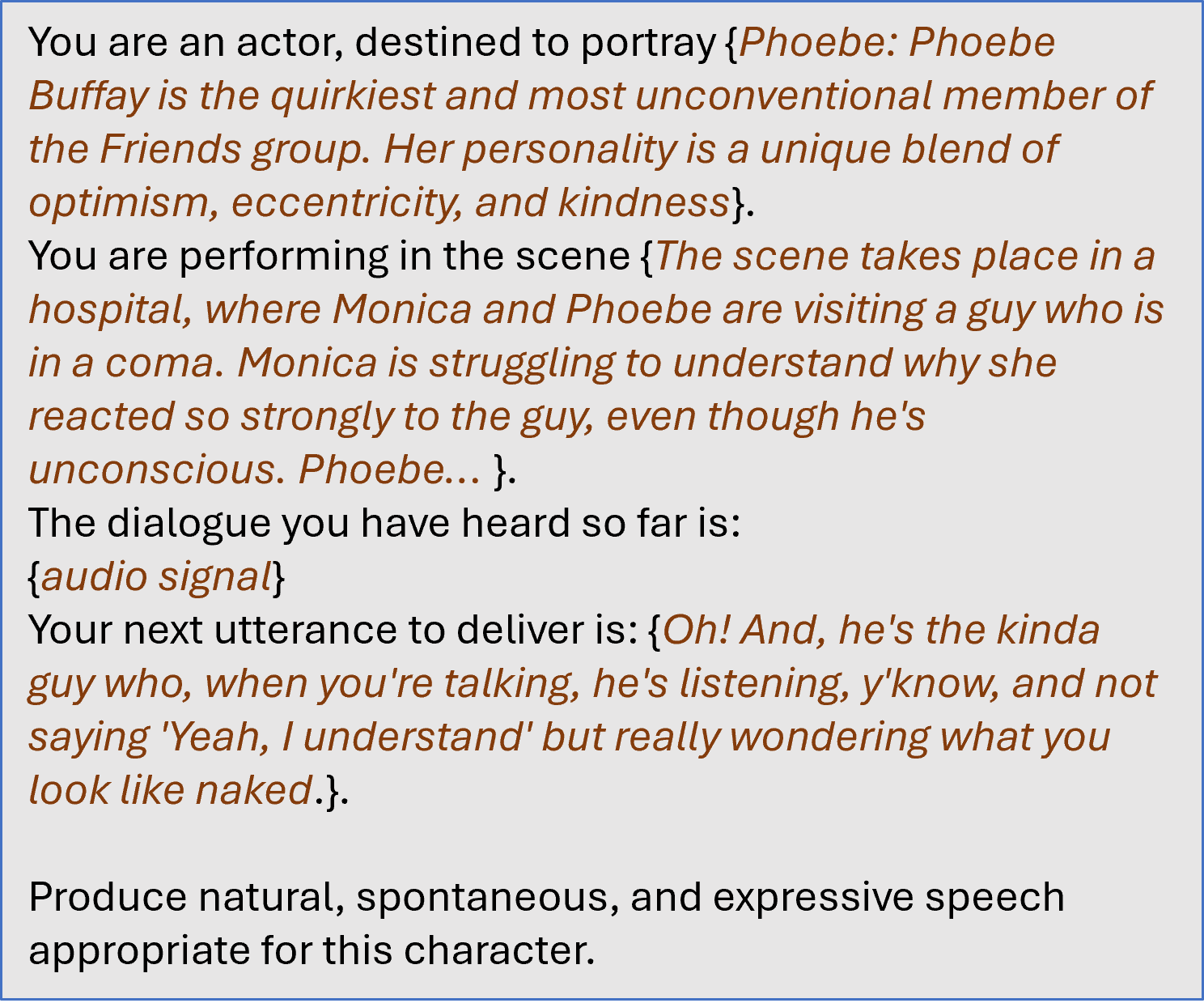}
  \caption{
  Qwen\_Omni Prompt for Speech Role-Playing. Black text: prompt template; Brownish-yellow: corresponding speech and text content.
  }
  \label{fig:bsl_qwen_prompt}
\end{figure}

\subsection{RP-MOS}
\label{appendix:EXP_metric}

This subjective evaluation assesses the model’s speech role-playing ability by comparing generated speech with reference (ground-truth) recordings. 
Participants listen to the generated speech and assign scores based on its similarity to the reference speech. 
Ten english speakers evaluated twelve rounds (six roles, with two sets per role), using randomly selected utterances across all model variants. 
All evaluators were provided with detailed guidelines and evaluation criteria prior to the assessment.

Each evaluator was compensated at a rate aligned with the average local hourly income, which we consider fair and appropriate given their country of residence and time commitment.

\paragraph{Guidelines and Evaluation Criteria}

Participants should consider the following aspects of emotional expression:

\begin{itemize}
    \item \textbf{Emotional Consistency}: Does the generated speech convey the same emotional tone as the reference speech?
    \item \textbf{Intensity Alignment}: Is the strength or intensity of the emotion comparable between the two speeches?
    \item \textbf{Naturalness and Realism}: Does the generated audio sound naturally expressive and believable, rather than artificial or flat?
    \item \textbf{Overall Impression}: Considering all the above factors, how similar is the emotional quality of the generated speech to the real reference?
    \item \textbf{Voice and Content Consistency}: If the voice is from different people or the text content differs from the reference, directly assign a score of 1.
\end{itemize}

\paragraph{Scoring Scale}

Use the following 5-point scale to rate each speech:

\begin{itemize}
    \item \textbf{5 – Identical}: The generated speech conveys the same emotion as the reference speech, with nearly identical intensity and expression.
    \item \textbf{4 – Very Similar}: The emotion is highly similar, with only minor differences in tone or intensity.
    \item \textbf{3 – Moderately Similar}: The overall emotion is recognizable but with noticeable differences in strength, tone, or expression.
    \item \textbf{2 – Weak Similarity}: The emotion type is somewhat related but largely inconsistent with the reference speech.
    \item \textbf{1 – No Similarity}: The generated audio conveys a completely different or unrecognizable emotion compared to the reference; If the voice seems to be from a different speaker, or if the text content differs from the reference, directly assign a score of 1.
\end{itemize}

\paragraph{Evaluation Procedure}

\begin{enumerate}
    \item Listen to each speech at least twice before rating.
    \item If the voice is from different people or the text content differs from the reference, directly assign a score of 1.
    \item Assign a single integer score (1–5) according to the criteria above.
    \item If uncertain, choose the score that best represents your overall impression.
\end{enumerate}

\subsection{ActorMind Implementation Detail}
\label{appendix:AM_tool}

\textbf{\textcolor{olive}{Eye Agent}} reads the preparatory descriptive content and retains it in memory. In practice, a textual memory of a few hundred words is sufficient.

\textbf{\textcolor{cyan}{Ear Agent}} Speech Emotion Captioning (SECAP) provides textual and intuitive emotional descriptions of target speech signals. SECAP \citep{secap} equips the \textcolor{black}{Ear Agent} with listening and emotion-recognition capabilities.

\textbf{\textcolor{violet}{Brain Agent}} is the central component of ActorMind. Following previous LLM role-playing works \citep{mmrole, rolellm}, we use LLama3 \citep{llama3} to perform emotional state reasoning, with the prompts illustrated in Figure~\ref{fig:AM_brain_pt}.

\begin{figure*}[htbp]
  \centering
  \includegraphics[width=\linewidth]{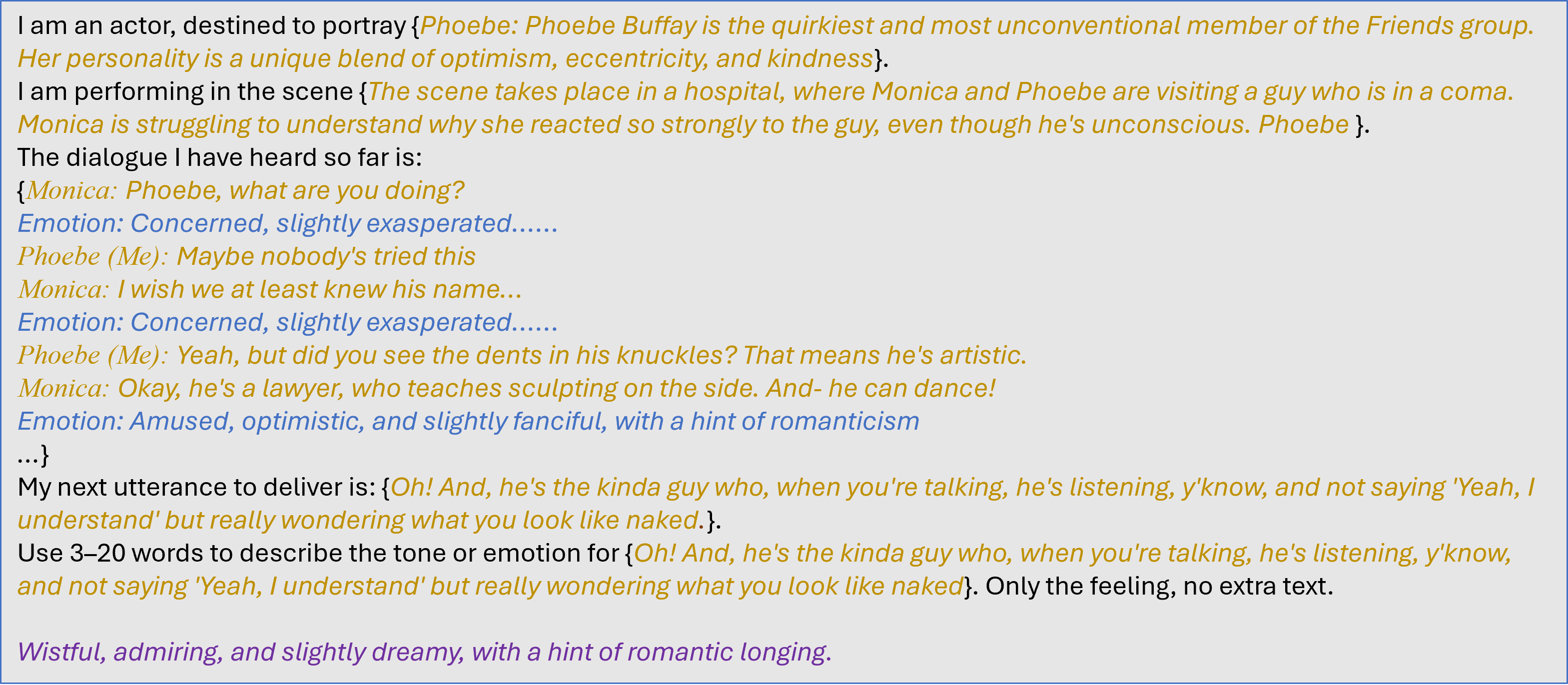}
  \caption{
     Prompt for the \textcolor{violet}{Brain Agent} used for role injection, contextual understanding, and emotion rendering. Here, \textcolor{black}{yellow content} represents what the \textcolor{olive}{Eye Agent} saw, \textcolor{black}{blue content} represents what the \textcolor{cyan}{Ear Agent} heard, and \textcolor{black}{purple content} represents what the \textcolor{violet}{Brain Agent} inferred. 
  }
  \label{fig:AM_brain_pt}
\end{figure*}

\textbf{\textcolor{orange}{Mouth Agent}} employs RAG to retrieve relevant context from a database for speech generation.
In ActorMind, for each role $R_k$, the database $Database_k$ is constructed from that role’s known speech utterances. 
Each entry contains the speech signal $U_x^s$ as content, with indices corresponding to emotional descriptions $E_x$ generated by SECAP \citep{secap}.  
During the retrieval phase, embeddings are computed using OpenAI's \textit{text-embedding-3-large}\footnote{\url{https://platform.openai.com/docs/models/embeddings}}.  
During the generation phase, we employ IndexTTS \footnote{\url{https://github.com/index-tts/index-tts}} \citep{indextts} as speech synthesizer, where the text prompt is the target line and the tone and emotion prompt is the retrieved speech.

\end{document}